\documentclass[a4paper, 10 pt, conference]{ieeeconf}  

\usepackage{amsmath,amssymb}
\usepackage[overload]{empheq}
\usepackage{graphicx}
\usepackage[export]{adjustbox}
\usepackage[dvips]{epsfig}
\usepackage{epstopdf}
\usepackage[table]{xcolor}
\usepackage{url}
\usepackage{subfig}
\usepackage{multirow}
\usepackage{siunitx}
\usepackage{eurosym}
\usepackage{enumerate}

\renewcommand\baselinestretch{0.945}

\newcommand{\quotes}[1]{``#1''}
\newcommand*\xbar[1]{%
  \hbox{%
    \vbox{%
      \hrule height 0.5pt 
      \kern0.5ex
      \hbox{%
        \kern-0.1em
        \ensuremath{#1}%
        \kern-0.1em
      }%
    }%
  }%
}

\def\sqrtexplained#1{%
  \begingroup
    \sbox0{$#1$}
    \def\underbrace##1_##2{##1}
    \sbox2{$#1$}
    \dimen0=\wd0 \advance\dimen0-\wd2
    \mathrlap{\sqrt{\phantom{\displaystyle#1}\kern\dimen0 }}
    \hphantom{\sqrt{\vphantom{\displaystyle#1}}}
  \endgroup
  #1}

\newcommand{\pushright}[1]{\ifmeasuring@#1\else\omit$\displaystyle#1$\ignorespaces\fi}

\IEEEoverridecommandlockouts                              

\title{\LARGE \bf Analysis and development of an automatic \textit{eCall} for motorcycles:\\ a one-class cepstrum approach}

\author{Simone Gelmini$^*$, Giulio Panzani and Sergio Savaresi
\thanks{The authors are with the Dipartimento di Elettronica, Informazione e Bioingegneria, Politecnico di Milano, Piazza Leonardo da Vinci 32, 20133 Milan, Italy. \newline 
$^*$Corresponding author: \texttt{\small{simone.gelmini@polimi.it}}. \newline This research is partly supported by the EU-sponsored project i-Hero.}
}%

\begin{document}

\maketitle
\thispagestyle{empty}
\pagestyle{empty}

\begin{abstract}
The automatic dial of an emergency call -- \textit{eCall} -- in response to a road accident is a feature that is gaining interest in the intelligent vehicle community. It indirectly increases the driving safety of road vehicles, but presents the technical challenge of developing an algorithm which triggers the emergency call only when needed, a non-trivial task for two-wheeled vehicles due to their complex dynamics. In the present work, we propose an \textit{eCall} algorithm that detects these anomalies in the data time series, thanks to the cepstral analysis. The main advantage of the proposed approach is the direct focus on the data dynamics, solving the limits of approaches based on the analysis of the instantaneous value of some signals combination. The algorithm is calibrated and tested against real driving data of ten different drivers, including seven real crash events, and performance are compared with known methods.
\end{abstract}

\section{Introduction}
According to the most recent data, the number of road traffic deaths continues to climb reaching 1.35 million in 2016, the 8th leading cause of death for people of all ages \cite{world2018global}. At least in Europe, this is partially due to the sharply increasing number of motorcycles -- and, more generally, powered two-wheelers (PTW) \cite{yannis2010road} -- with their 34-fold higher risk of death in a crash than the other motor vehicles users \cite{lin2009review}. Motorcyclist fatalities already account for almost $11\%$ of the total in Europe \cite{world2018global}.

One possible solution to the problem is the development of safety-related vehicle dynamics control. In spite of the effort and the improved vehicles, there is still more to do for reducing this negative trend. To this end, as second pillar, regulatory agencies have started focusing on the improvement of the rescuing operations that follow an accident, because \quotes{the immediate transport to an appropriate trauma center is one of the essential steps  in the early treatment of polytraumatized patients} \cite{krettek1998management}. More specifically, the EU has promoted an initiative, \textit{eCall}, aiming to automatically dial an emergency call when a road accident happens, investing in a large project known as Infrastructure Harmonised \textit{eCall} European Pilot (i-Heero). The aim of the \textit{eCall} is to detect potentially dangerous events whose dynamics are compatible to a crash, followed by a long enough time interval of inaction, the most recurrent pattern in road accidents. This project has already been transformed in a set of rules, leading to the adoption of a standard equipment to be installed on all cars in EU since the end of 2018, \cite{uhlemann2015introducing}, and an extension for motorcycles is under development. This work does not present the main results obtained in the i-Heero project, but it is a theoretical and research-oriented extension of such results.

As far as crash detection is concerned, such event essentially takes place in the ground plane in four-wheeled vehicles. Thus, its detection can be obtained by monitoring abnormal longitudinal decelerations, see \textit{e.g.}, \cite{gu2016twitter}. Instead, the crash becomes more difficult to be defined in a unique way when moving to two-wheeled vehicles, since their motion is strongly affected by the roll angle dynamics. The few works found in literature on this topic can be clustered in two main categories. The first one can be labelled as \textit{threshold-based}: since PTW crashes are most often paired with the fall of the vehicles, static threshold methods like \cite{boubezoul2013simple,giovannini2014development,bellati2006preliminary} aim to detect the occurrence of a crash monitoring when the motorcycles are suspiciously not upright. These methods are usually the result of a significant analysis of the falling dynamics, and the detection is generally obtained when some post-processed signals exceed the given thresholds. The second approach is called \textit{statistical-based}: as presented in \cite{attal2014powered,attal2015powered,vlahogianni2013critical}, these methods assume a crash to be an anomaly in the data distribution, and the detection is realized using statistical tools like outliers-detection algorithms. These methods are also capable of detecting crashes that happen when the vehicles remain upright, events not detected by threshold-based ones.

In all the published works, the dynamics of the recorded signals -- which are usually some inertial measurements of the vehicle -- are not explicitly accounted. However, in the analysis of dynamic systems like PTWs, the role of time is central. In fact, the analysis of structured-time signals allows to exploit the intrinsic correlations between them. To this purpose, a frequency-domain analysis is possible, which allows inferring on the underlying physical phenomenon by inspecting the most discriminating harmonic ranges by means of efficient mathematical tools, like cepstrum \cite{decock2002phd}. 

In this paper we present a novel way to detect motorcycle crash events through the dynamical analysis of the vehicle's motion. To do this, recorded data are remapped in the cepstrum domain and then monitored, detecting an anomaly when the driving dynamics differs significantly from its regular trend. To our best knowledge, this is the first work where cepstrum-based signal processing is used for driving anomaly detection.

\section{Problem statement and experimental setup} \label{sec:set_up}
In this paper, an automatic \textit{eCall} algorithm for two-wheeled vehicles is presented. The goal of the proposed contribution is to analyze the underlying driving dynamics and searching for patterns not compatible with the regular ride of the vehicle, aiming to detect any event in which drivers need to be rescued (\textit{e.g.}, after falls or collisions). Due to potentially impelling medical conditions, the detection needs to be performed shortly after the crash to provide a prompt assistance. Furthermore, the algorithm must detect all the anomalies while limiting the social and economical costs of triggering emergency calls unnecessarily.

\begin{figure}[thpb]
 \centering
 \includegraphics[trim={100 140 100 150},clip,width=0.35\textwidth]{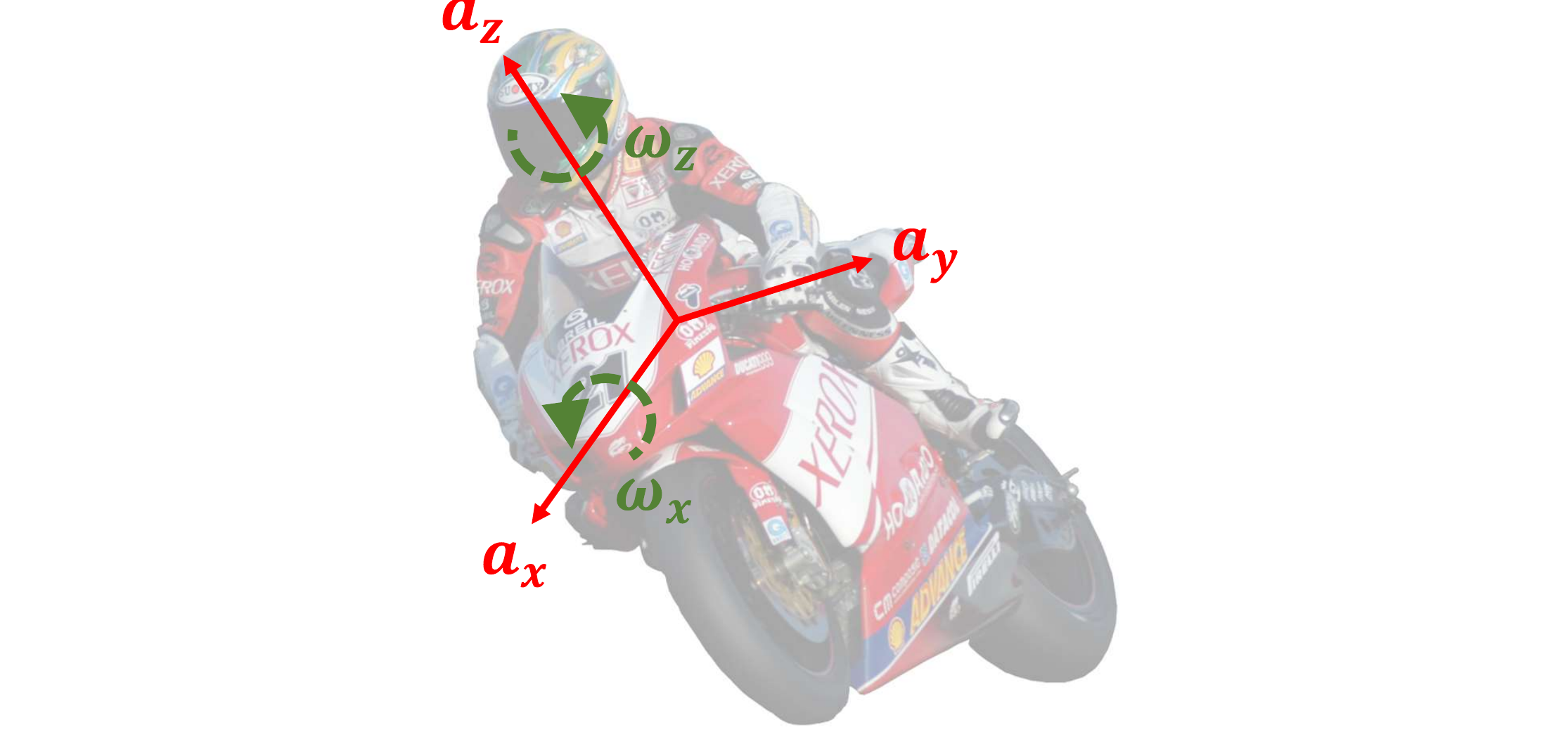}
 \caption{An overview of the standard reference frame.}
 \label{fig:setup}
\end{figure}

In order to widespread the use even in outdated motorcycles, the algorithm is designed so to use a minimal sensor setup. In contrast to the sophisticated sensor layouts like those used in other contributions, \textit{e.g.}, \cite{boubezoul2013simple,vlahogianni2013critical}, the proposed \textit{eCall} trigger is obtained only using data of a 5 degrees of freedom intertial measurement unit (IMU) -- aligned with respect to the standard reference frame, as depicted in Fig. \ref{fig:setup} -- which embeds a triaxial accelerations vector and two angular velocities, the roll-rate ($\omega_x$) and yaw-rate ($\omega_z$). This is the minimal sensor configuration already employed in recent motorcycles for estimating the lean angle \cite{boniolo2009roll} and which could be easily replicated using the modern, flexible, cost-effective, and easy to install telematic e-Boxes \cite{gelmini2018selfcalibration}. 
In the following, the available vehicle speed is presented only for monitoring purposes and for testing one of the algorithms already presented in literature, but it is not used for the proposed approach.

The available data have been collected during an experimental campaign involving ten professional riders during on-track tests, yielding more than eighty hours of recordings, including seven different crash events (listed in Table \ref{tab:length_crash}).
\scriptsize
\begin{table}[thpb]
\centering
\caption{Summary of the recorded crash types and their duration.}
\begin{tabular}{cc}
\hline	
\textbf{Event} & \textbf{Crash duration} $\mathrm{[s]}$\\
\hline
\hline
Front Lowside I 	& $10$\\  
Front Lowside II 	& $8$\\  
Cornering Lowside 	& $10$\\	  
Highside I 			& $7$\\  
Highside II 		& $9$\\
Sliding I 			& $6$\\
Sliding II 			& $14$\\
\hline
\end{tabular}
\label{tab:length_crash}		
\end{table}
\normalsize
\section{Related works} \label{sec:related_works}
\subsection{Threshold-based algorithms}
Threshold-based algorithms aim to detect anomalous riding patterns by means of a combination of signal processing and static thresholds. These algorithms generally reflect the intuitive and simple idea of crashes, mainly related to the motorcycle impact and/or its fall to the ground, that result in \quotes{extremely large} signals that overcome normal values.

For instance, the work discussed in \cite{boubezoul2013simple} detects the occurrence of a fall when the driver is close to experience a nominal free-fall condition. Given the acceleration $\|a(t)\|=\sqrt{a_x(t)^2+a_y(t)^2+a_z(t)^2}$ and angular rate $\|\omega(t)\|=\sqrt{\omega_x(t)^2+\omega_y(t)^2+\omega_z(t)^2}$ norms, an emergency call is fired when $\|a(t)\|<\gamma_{a}=0.5g$, in which $g=9.8056\ \mathrm{m/s^2}$ is gravity, and $\|\omega(t)\|>\gamma_{\omega}=2\ \mathrm{rad/s}$.

We test this intuitive and easy to replicate algorithm against our set of data. However, since the pitch-rate ($\omega_y$) is not available in our setup, the angular rates norm is computed only on axes $x-z$, $\|\omega(t)\|_{x-z}=\sqrt{\omega_x(t)^2+\omega_z(t)^2}$.
\begin{figure}[thpb]
 \centering
 \includegraphics[width=0.42\textwidth]{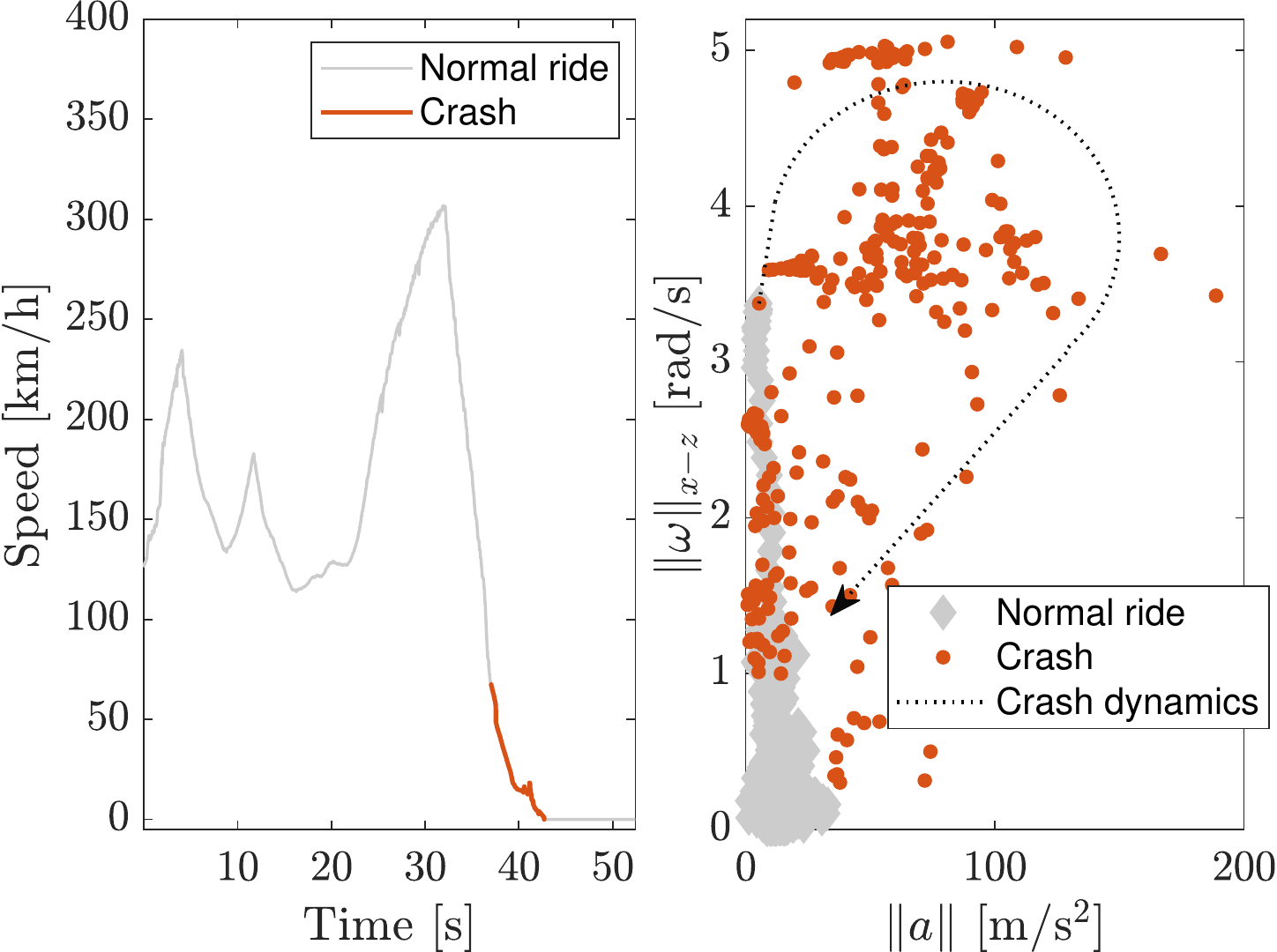}
 \caption{Analysis of the pattern of a falling motorcycle for the \textit{Front lowside I} crash.}
 \label{fig:crash_dynamics}
\end{figure}

In Fig. \ref{fig:crash_dynamics} the dynamics of the \textit{Front lowside I} is analyzed: during the fall, the angular rate norm grows and the acceleration one floats towards zero; at the impact, the two norms reach their peaks, with the vehicle bouncing on the ground until it settles. Even with a missing degree of freedom, the pattern still resembles the one discussed in the original paper. However, for a fair comparison, the algorithm has been recalibrated, ensuring the minimal tuning that guarantees the detection of all the crashes, obtaining $\gamma_{a}=7.86\ \mathrm{m/s^2}$ and $\gamma_{\omega}=1.77\ \mathrm{rad/s}$. Under this parametrization, the algorithm is yet able to detect the falls opportunely. 

\begin{figure}[thpb]
	\centering
	\includegraphics[width=0.42\textwidth]{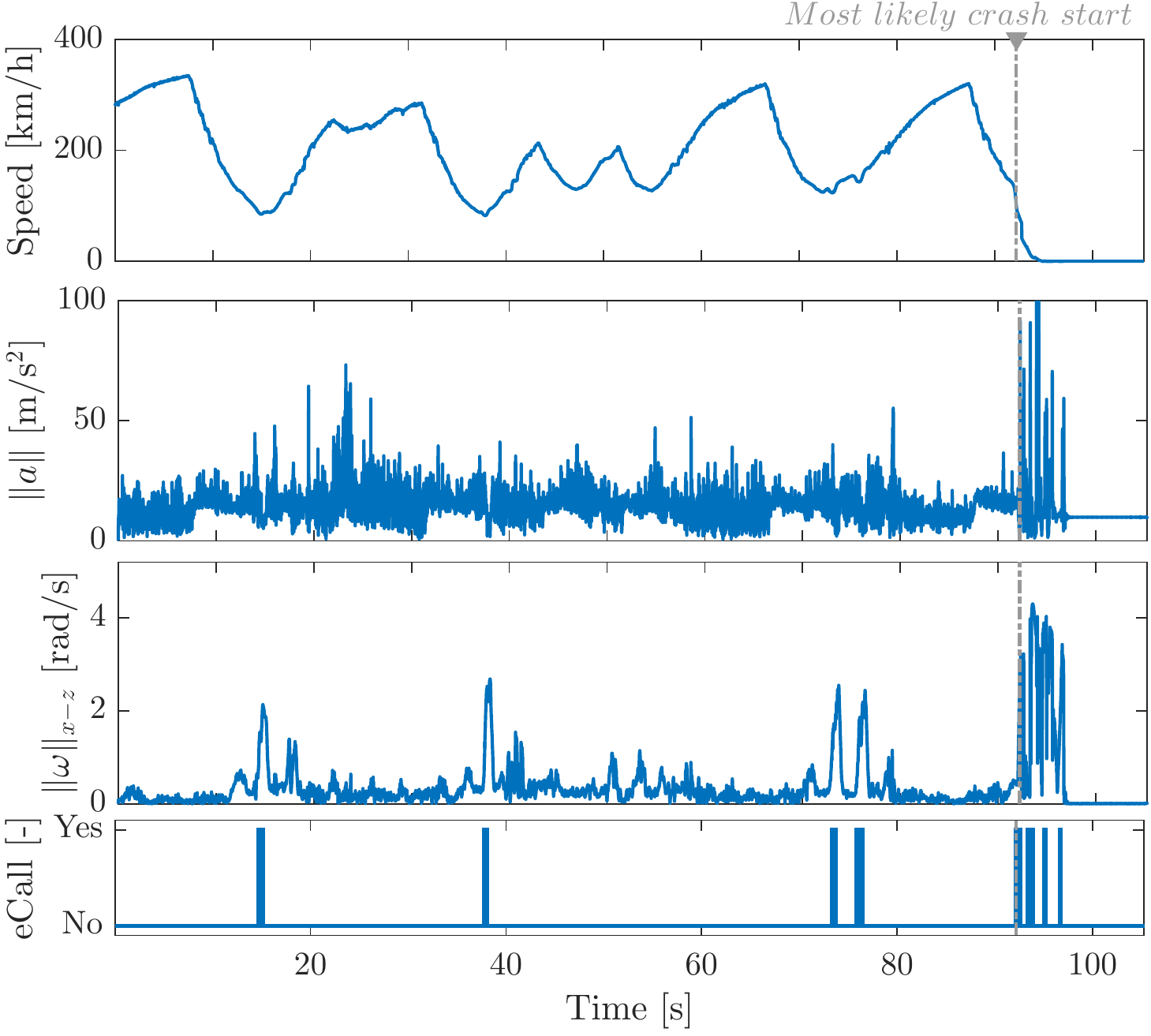}
	\caption{An example in which the threshold-based algorithm generates many false positives before the actual \textit{Cornering lowside} crash.}
	\label{fig:threshold_based_fall_false_positives}
\end{figure}

The main disadvantage of the threshold-based approach is that it results in a significant number of false-positive, triggering an \textit{eCall} unnecessarily, as shown in Fig. \ref{fig:threshold_based_fall_false_positives}: besides the correct crash trigger, several activations are visible during the normal driving scenario. 

\begin{figure}[thpb]
	\centering
	\includegraphics[trim={45 45 45 45},clip,width=0.42\textwidth]{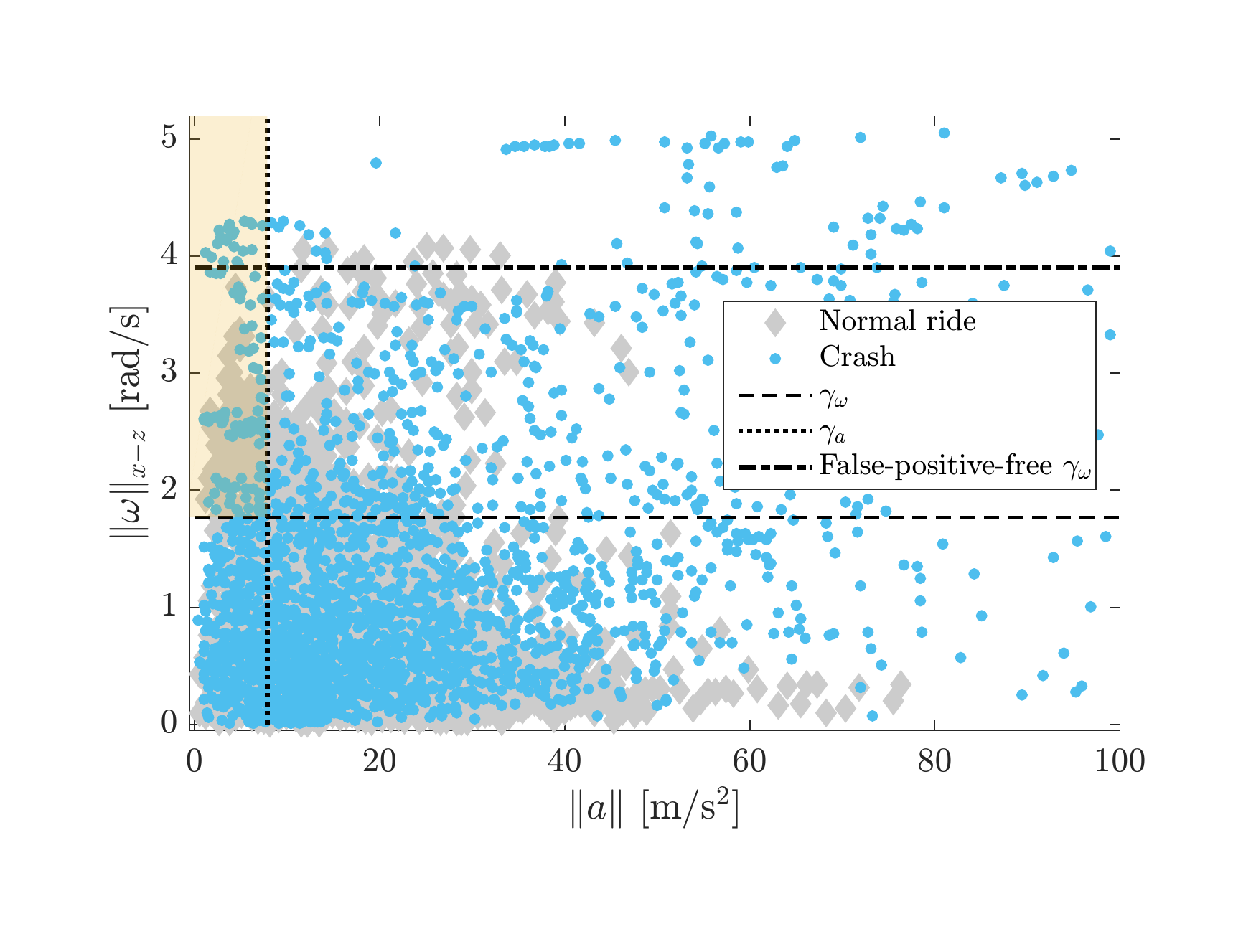}
	\caption{Analysis of the detected crash instances vs the regular riding distribution for the \textit{Cornering lowside} crash.}
	\label{fig:sensitivity_threshold_based}
\end{figure}

A better understanding is provided by Fig. \ref{fig:sensitivity_threshold_based}, which depicts the \textit{eCall} triggering area (\textit{i.e.}, an emergency is detected if a sample falls within the shadowed area): the normal driving data within the shadowed area are responsible for the false positive activations. The inspection of the same plot suggests that a different threshold tuning (\textit{e.g.}, increasing the angular acceleration norm threshold, portrayed in the illustration as \textit{False-positive-free} $\gamma_{\omega}$) minimizes the number of incorrect activations, since no normal riding data would fall in the \textit{eCall} triggering one. This apparently positive result comes with the unacceptable consequence of delayed or even missed crash detections (false negative events), ignoring potentially dangerous situations that must be detected.

As a result, static threshold methods do not seem a viable solution in real \textit{eCall} applications since false positives cannot be systematically eliminated. A possible workaround is to equip the motorcycle with an interface (\textit{e.g.} a flashing light button) which alerts the drivers when the algorithm is triggered, allowing them to interrupt the emergency call if unnecessary. However, this solution might be still dangerous as it would distract and annoy the motorcyclist, especially in case of frequently repeated false positives.

\subsection{Statistical-based algorithms}
Beside threshold-based algorithms, the scientific literature proposes a different approach to the problem, based on statistical analysis. These algorithms, see, \textit{e.g.}, \cite{attal2014powered,attal2015powered,vlahogianni2014detecting}, assume that accidents are anomalies with respect to the nominal data distribution, identifying them through outlier detection methods. In such framework, an accident is the sample that does not statistically belong to the main cluster, for a certain significance level. Statistical-based crash detection algorithms can be considered an evolution of the threshold-based ones since the former analyzes the data through synthetic, independently distributed, features, whereas the latter analyzes the data as a multivariate distribution, exploiting the intrinsic correlations not accounted previously.

\begin{figure}[thpb]
 	\centering
 	\includegraphics[width=0.42\textwidth]{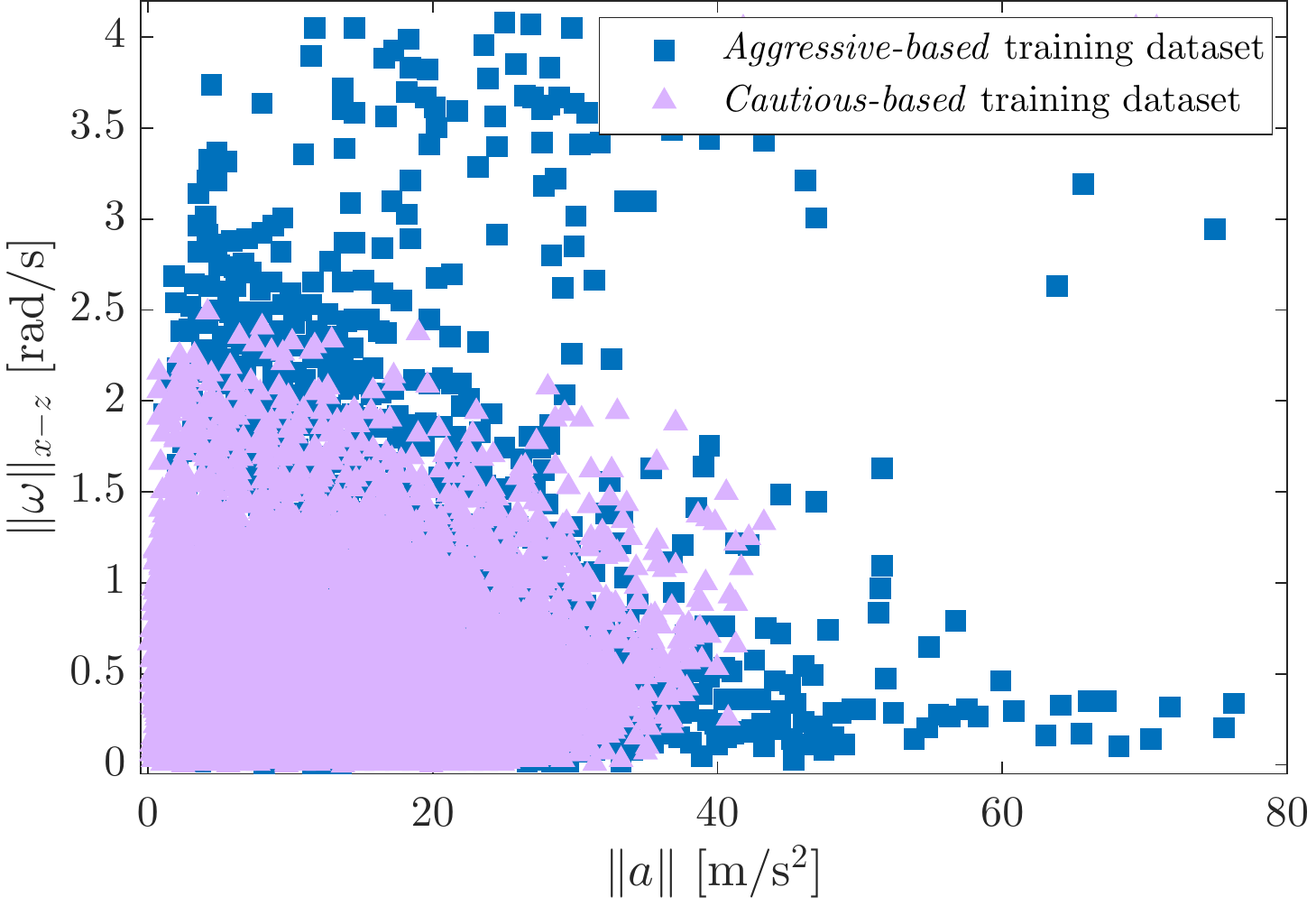}
 	\caption{The effects of the driving style on the data distribution for the \textit{Cornering lowside} crash: the shape and amplitude change when considering cautious or aggressive drivers.}
 	\label{fig:sensitivity_statistics_based_reduced}
\end{figure}

In \cite{vlahogianni2014detecting}, Vlahogianni and authors propose to detect the anomalies (\textit{i.e.}, the accidents) by means of the Mahalanobis distance, which provides a quantitative metric for assessing how far a sample vector is from a given (possibly multivariate) distribution. Denoting the sampled data at each time instant with $l(t)\in \mathbb{R}^{p\times 1}$ (with $p$ the number of signals analyzed), such distance is formulated as  
\begin{equation}
d_{Mah}(t)=\sqrt{\left(l(t)-\hat{\mu} \right)'\hat{S}^{-1} \left(l(t)-\hat{\mu} \right)},
\label{eq:Maha}
\end{equation}
in which $\hat{\mu}$ and $\hat{S}$ are the sampled mean and covariance matrix computed on the training dataset. A sampled data $l(t)$ represents an anomaly if the Mahalanobis distance in (\ref{eq:Maha}) exceed a certain threshold, $d_{Mah}(t) > \gamma_{Mah}$, triggering an emergency call. The advantage of using such an approach is twofold: on one side, the shape of the normal driving data distribution is taken into account; on the other, given a statistical significance level $\alpha$ (in this application, as in the original paper, $\alpha=0.05$), the threshold $\gamma_{Mah}$ is learned from data since the given Mahalanobis distance can be approximated to an $F-$distributed variable
\begin{equation}
d_{Mah} \sim \frac{p(n-1)(n+1)}{n(n-p)} F_{p,n-p},
\label{eq:fisher}
\end{equation}
in which $n$ is the multivariate space dimension (\textit{i.e.}, the dataset's number of sampled vectors). Moreover, limited by the reduced sensor configuration of the present work, the benchmark results presented in the following refer to model C in \cite{vlahogianni2014detecting}, whose sample data include the vehicle speed and all the available inertial measurements. Analogous results are obtained for the smaller subset of sensors (\textit{i.e.}, vehicle speed and longitudinal acceleration), accounted by model B in \cite{vlahogianni2014detecting}.
\begin{figure}[thpb]
	\centering
	\includegraphics[width=0.42\textwidth]{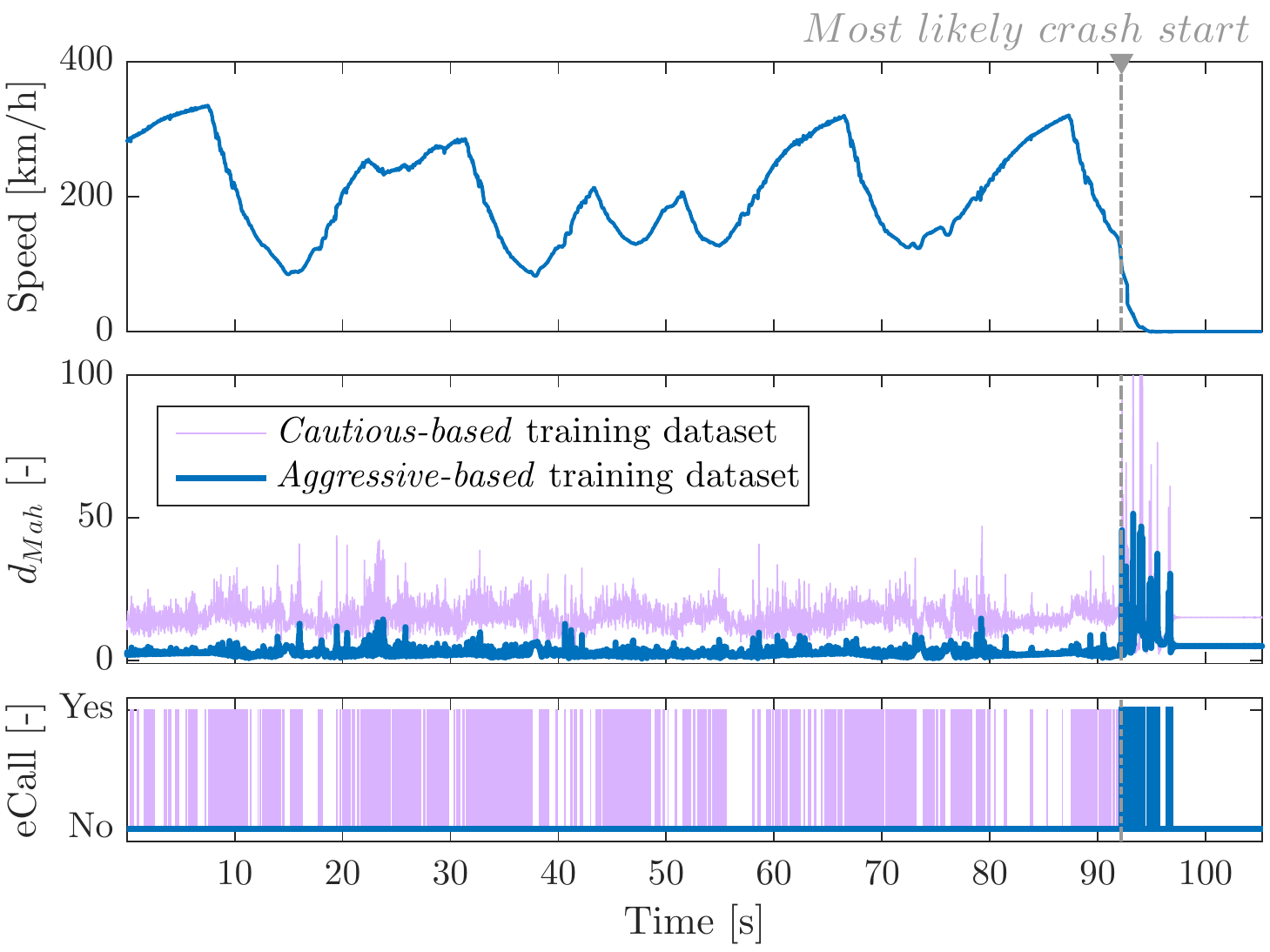}
	\caption{Performance analysis of the statistical-based approach trained on two different driving styles datasets for the \textit{Cornering lowside} crash.}
	\label{fig:statistical_based_model_all}
\end{figure}
Figure \ref{fig:statistical_based_model_all} shows the detection results of the statistical-based method, on the same crash previously presented in Fig. \ref{fig:threshold_based_fall_false_positives}. The analysis of the middle plot, where the computed Mahalanobis distance is depicted, highlights the high sensitivity of this approach with respect to the training datasets: as qualitatively shown in in Fig. \ref{fig:sensitivity_statistics_based_reduced}, the two distributions are characterized by two different riding styles, which result in different sampled mean and covariance values, both part of the distance definition \eqref{eq:Maha}. The crash is detected correctly in both the two situations, but the \textit{cautious-based} trained algorithm results in a higher number of false positives.

Although the statistical-based approach proves to be more robust when the training distribution well approximates the underlying one (in fact, no false positives are generated in the \textit{aggressive-based} case), it is not likely to be able to define a reference distribution for all the drivers. Besides, even if a proper choice of the training dataset can significantly reduce the number of false activations, statistical-based crash detection approaches still suffer of another drawback, namely the high variability of the resulting classification signal (\textit{i.e.}, the frequent ON/OFF switches), which requires to post-process the output (\textit{e.g.}, through some debounce logic), ensuring the consistency of the detected events, eventually delaying or missing the detection. Furthermore, the method cannot detect accidents whose dynamics lies within the main distribution, implicitly assuming that a driver is always able to control the vehicle under the same dynamic conditions.

These drawbacks are inherent of the nature of the two approaches, as they both classify each incoming data sample independently from its past values. This motivates the development of a different algorithm, capable of accounting for the signal dynamics. Such approach is the main core of the present work and it is presented in the following section.

\section{One-class cepstrum classification} \label{sec:one_class_cepstrum}
\subsection{Dynamical analysis}
\begin{figure*}[thpb]
\centering
\subfloat[Normal ride]{\begin{tabular}[b]{c}
					   		\includegraphics[width=0.4\textwidth]{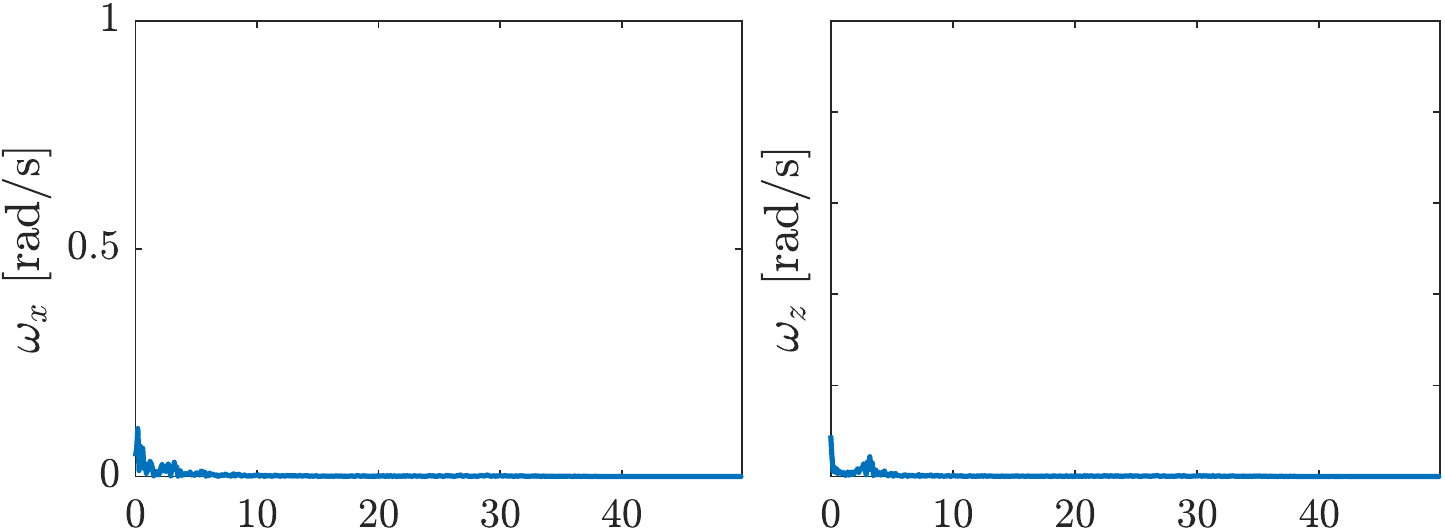}\\
					   		\includegraphics[width=0.4\textwidth]{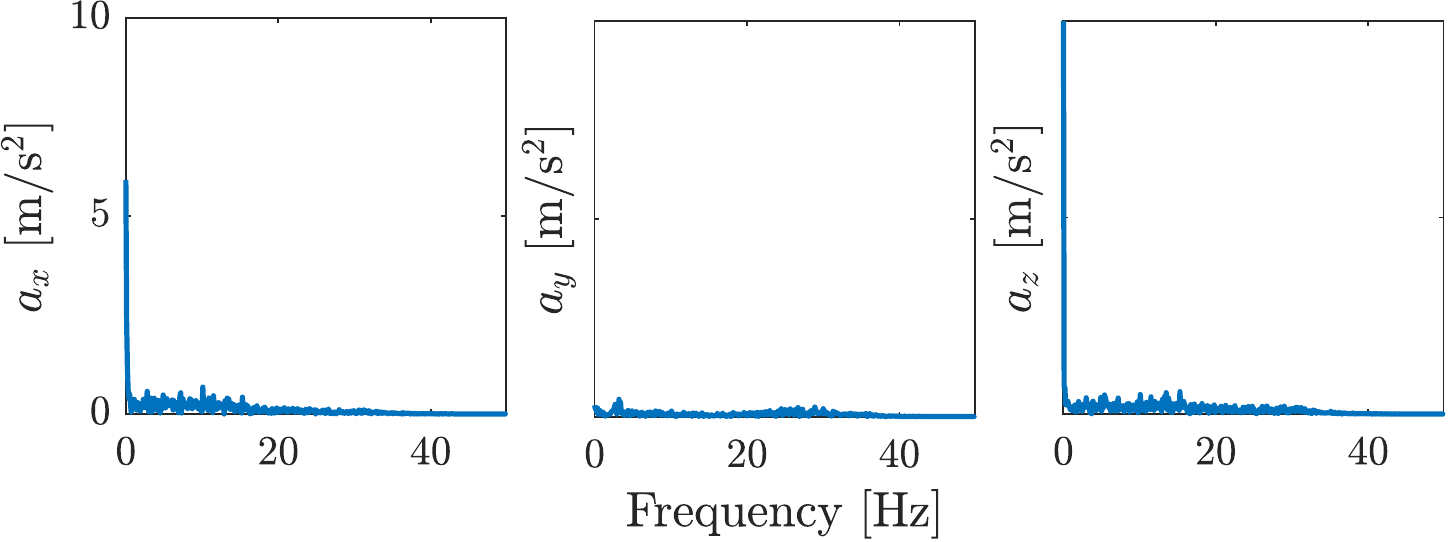}
				       \end{tabular}}
\subfloat[Crash dynamics]{\begin{tabular}[b]{c}
							\includegraphics[width=0.4\textwidth]{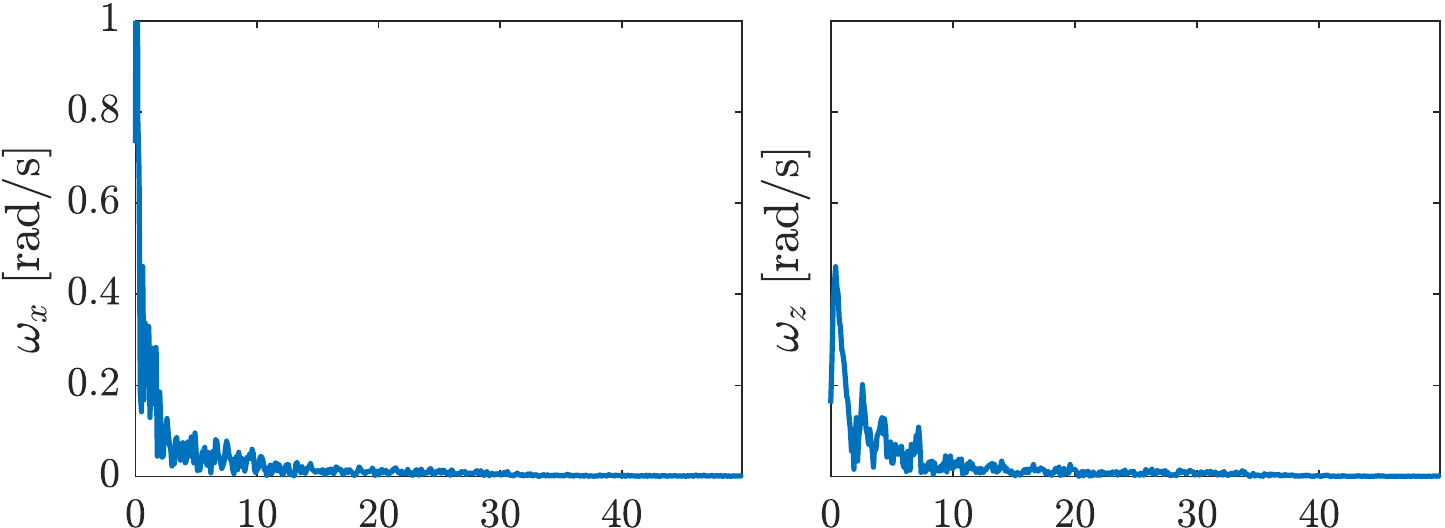}\\
							\includegraphics[width=0.4\textwidth]{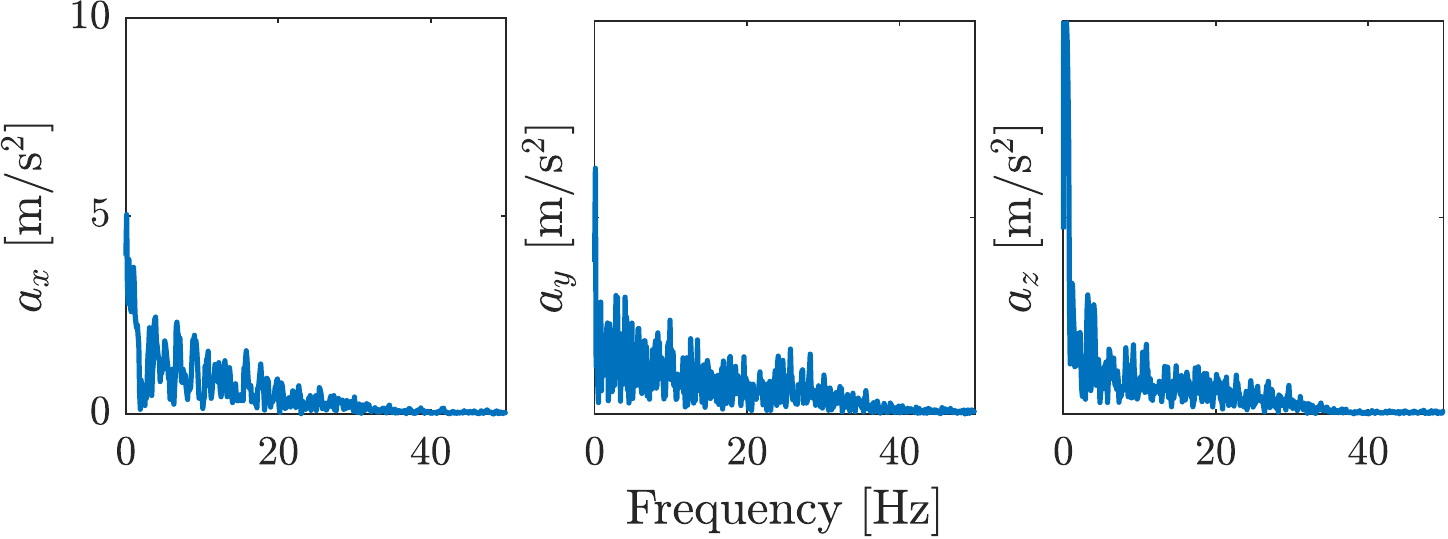}
						   \end{tabular}}
\caption{Comparison of the spectra in a window during the normal ride and the crash dynamics: most of the spectra harmonics are highly excited during the crash event, compared to the normal ride of the vehicle.}
\label{fig:spectra}
\end{figure*}

To improve the detection performance, the crash dynamics is exploited. Data are analyzed in the frequency-domain, as illustrated in Fig. \ref{fig:spectra}. Spectra are computed through a sliding window, before and during the crash event. As shown, the motorcycle does not significantly influence harmonics higher than the vehicle dynamics range (\textit{i.e.}, greater than $10\ \mathrm{Hz}$) during the normal ride. On the contrary, during the crash event, the whole spectra are excited due to the sequence of impacts and rotations. Since all the harmonics are excited, a simple classification approach based on the intensity of a high-pass filtered version of the signals would not extract enough information to perform a robust detection. Thus, to properly detect a crash, it is paramount to inspect all the inertial signals' spectrum, searching for variations in the frequency-domain which are not compatible with the regular ride of the vehicle.

\subsection{Cepstrum and related concepts}
To capture anomalies analyzing the data dynamics, we propose to use the power cepstrum, a mathematical tool introduced in \cite{bogert1963quefrency} as a better alternative of the autocorrelation function. As analyzed in \cite{kalpakis2001distance}, cepstrum is an effective tool for pattern recognition in time-structured problems as it has proved to have a higher discriminating power, though avoiding the curse of dimensionality.

The power cepstrum is defined as \quotes{the power spectrum of the logarithm of the power spectrum}. Assuming the inertial signals measured from the vehicle to be second-order quasi-stationary during the normal ride \cite{rahbar2005frequency}, within a given window each logged signal can be considered a stationary stochastic process $s(t)$ with spectrum $\Phi_s$. The power cepstrum can be computed as the inverse Fourier transform of the logarithm of the spectrum $\Phi_s$ \cite{decock2002phd}, as
\begin{equation}
\begin{split}
c_{s}(k)&=\frac{1}{2\pi} \int_0^{2\pi} \log \left(\Phi_{s}\left(e^{i\theta} \right)\right)e^{ik\theta}d\theta.
\end{split}
\label{cepstrum_definition}
\end{equation}

Furthermore, cepstrum is known to be a homomorphic system \cite{decock2002phd}: homomorphic systems are the ones in which nonlinear relationships could be converted into linear in their transform domains. For this reason, the cepstrum computed on a stationary stochastic process $v(t)$, obtained through the convolution of two stationary stochastic processes $v(t)=s_1(t)*s_2(t)=\int_{-\infty}^{+\infty}s_1(\tau)s_2(t-\tau)d\tau$, corresponds to the sum of the cepstrum computed on $s_1(t)$ and $s_2(t)$ separately
\begin{equation}
\begin{split}
c_v(k)&=\mathcal{F}^{-1}\left(\log \left(\Phi_v\right)\right)\\
&=\mathcal{F}^{-1}\left(\log\left(\Phi_{s_1}\right)\right)+\mathcal{F}^{-1}\left(\log\left(\Phi_{s_2}\right)\right)\\
&=c_{s_1}(k)+c_{s_2}(k).
\end{split}
\end{equation}

Thanks to the cepstrum's homomorphic property, we propose to extend the definition of the so-called Martin distance \cite{martin2000metric} -- so far used for univariate time series clustering and classification problems \cite{lauwers2017time} -- in order to account, in one metric, multiple time series of the same operating mode (\textit{i.e.}, same riding condition). Thus, assuming to record $p$ signals in two operating modes (\textit{e.g.}, $\mathbf{g}_a$ and $\mathbf{g}_b$), the distance becomes
\begin{equation}
d_{cep}(\mathbf{g}_a,\mathbf{g}_b)=\sqrtexplained{\sum_{k=0}^{\infty}k \Bigg| \sum_{j=1}^{p} c_{g_{a_j}}(k)-c_{g_{b_j}}(k) \Bigg| ^2},
\label{martin_distance_one_class}
\end{equation}
in which $r(k)=\sum_{j=1}^{p}c_{g_{a_j}}(k)-c_{g_{b_j}}(k)$ is the sum of the cepstral mismatch between the two operating modes, for all the $p$ features, for a given order $k$.

\subsection{Detection algorithm}
\begin{figure}[thpb]
 \centering
 \includegraphics[trim={220 130 220 135},clip,width=0.45\textwidth]{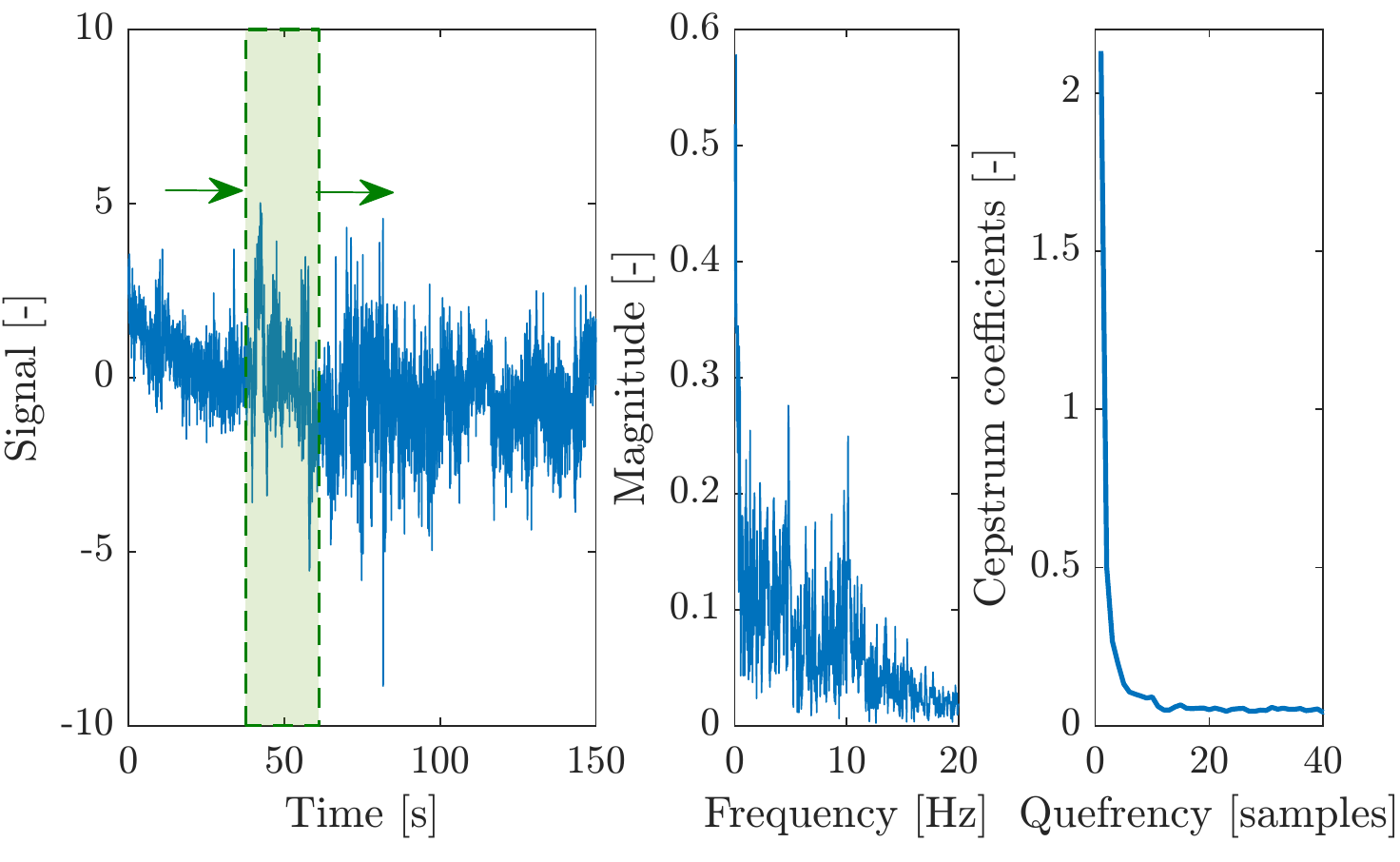}
 \caption{The process for calculating the cepstrum: the coefficients are evaluated on the spectrum computed on the windowed signal.}
 \label{fig:classifying}
\end{figure}

The proposed method automatically detects an accident through the measured inertial signals, when the underlying dynamics does not reflect the regular ride of the vehicle, by means of the homomorphic property of cepstrum. To this purpose, at any time instant $t$, a stream of data $\mathbf{s}\in\mathbb{R}^{p\times m}$ is given, in which $p$ is the number of signals recorded (\textit{e.g.}, in our case $p=5$), while $m=wf_s$ is the size of a sliding window buffering the last $m$ samples of each signal, in which $w$ is the window size in seconds and $f_s$ the sampling frequency (in this work, $f_s=100\ \mathrm{Hz}$):
\begin{equation}
\mathbf{s}(t)=\begin{bmatrix}s_{1}(t) & s_{1}(t-1) & \dots & s_{1}(t-m+1)\\
 \vdots & \vdots & \vdots & \vdots \\
s_{p}(t) & s_{p}(t-1) & \dots & s_{p}(t-m+1)
\end{bmatrix}.
\end{equation}
The cepstrum of each signal $s_j$ (with $j=1,\dots,p$) is computed as:
\begin{itemize}
\item first, the spectra of each time series $\Phi_{s_{j}}(t)=\frac{1}{m}\big| S_{j} \big|^2$ is evaluated, where $S_{j}$ is the periodogram computed on the portion of signal $s_{j}(t-m,\dots,t)$, obtained by means of the Fast Fourier Transform (FFT);
\item  then, the cepstrum coefficients of each signal $s_j$ (with $j=1,\dots,p$) are evaluated with the Inverse Fast Fourier Transform (IFFT) of the logarithm of the spectrum $\Phi_{s_{j}}$, or $c_{s_{j}}(k)=\mathrm{IFFT}\left( \log \left(\Phi_{s_{j}}(t)\right)\right)$, with $c_{s_{j}}\in\mathbb{R}^{1\times m}$.
\end{itemize}
The overall process is summarized in the example in Fig. \ref{fig:classifying}.

To detect the amplitude growth for all the signals' spectrum, the computed cepstrum is compared with a reference time series, whose cepstrum coefficients $\mathbf{o}\in\mathbb{R}^{p\times m}$ are all set to zero, representing a stream of data composed of constant signals. The chosen reference time series is independent from the driving style, making the analysis objective and robust for different drivers. Thus, the measured cepstrum coefficients $\mathbf{s}(t)$ are compared to the reference ones $\mathbf{o}$ thanks to the distance in \eqref{martin_distance_one_class}:
\begin{equation}
d_{cep}(\mathbf{s}(t),\mathbf{o}) =\sqrt{\sum_{k=0}^{\infty}k \Bigg| \sum_{j=1}^{p} c_{s_{j}}(k)-c_{o_{j}}(k) \Bigg| ^2}.
\label{martin_distance_one_class_application}
\end{equation}
An anomaly is detected when (\ref{martin_distance_one_class_application}) exceeds a tuned threshold $\gamma_{cep}$, meaning that the spectra have grown significantly with respect to the standard driving condition.

\section{Experimental results} \label{sec:results}
In this section, preliminary results on data recorded during an experimental campaign aiming to calibrate and validate the proposed method are discussed. Test data include seven real crashes and more than eighty hours of standard driving, in which no accidents are reported. First of all, the size of the buffering window $w$ is chosen such that the buffer is sufficiently large so to capture the crash event almost entirely. As reported in Table \ref{tab:length_crash}, except for one case, all the crash dynamics last less or equal ten seconds, thus the window length is set to $w=10\ \mathrm{s}$.

\begin{figure}[thpb]
	\centering
	\includegraphics[width=0.42\textwidth]{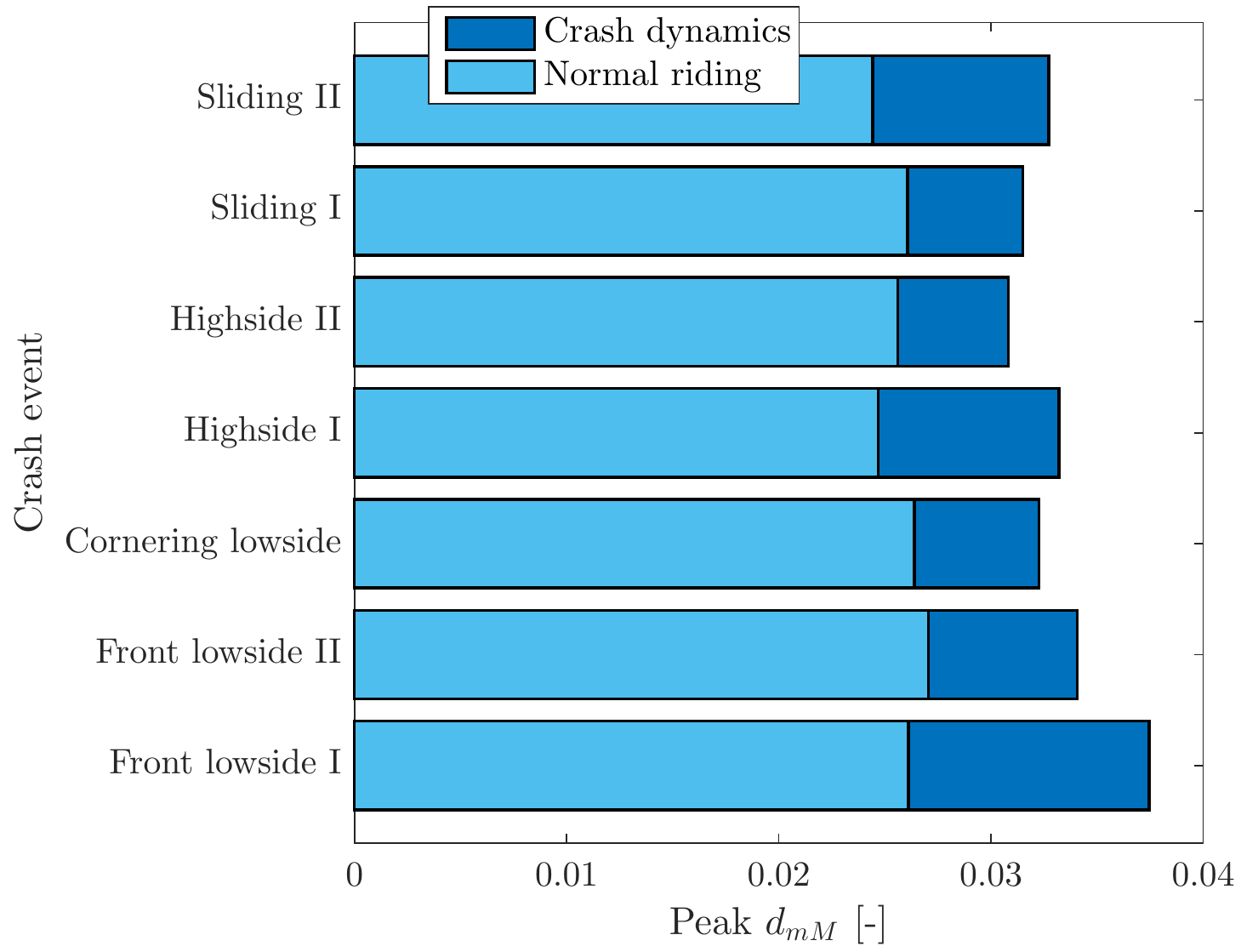}
	\caption{Analysis of the maximum value of the modified Martin distance $d_{cep}$ for both the normal and crash dynamics.}
	\label{fig:sensitivity_threshold_cepstrum}
\end{figure}

\begin{figure}[thpb]
	\centering
	\includegraphics[width=0.42\textwidth]{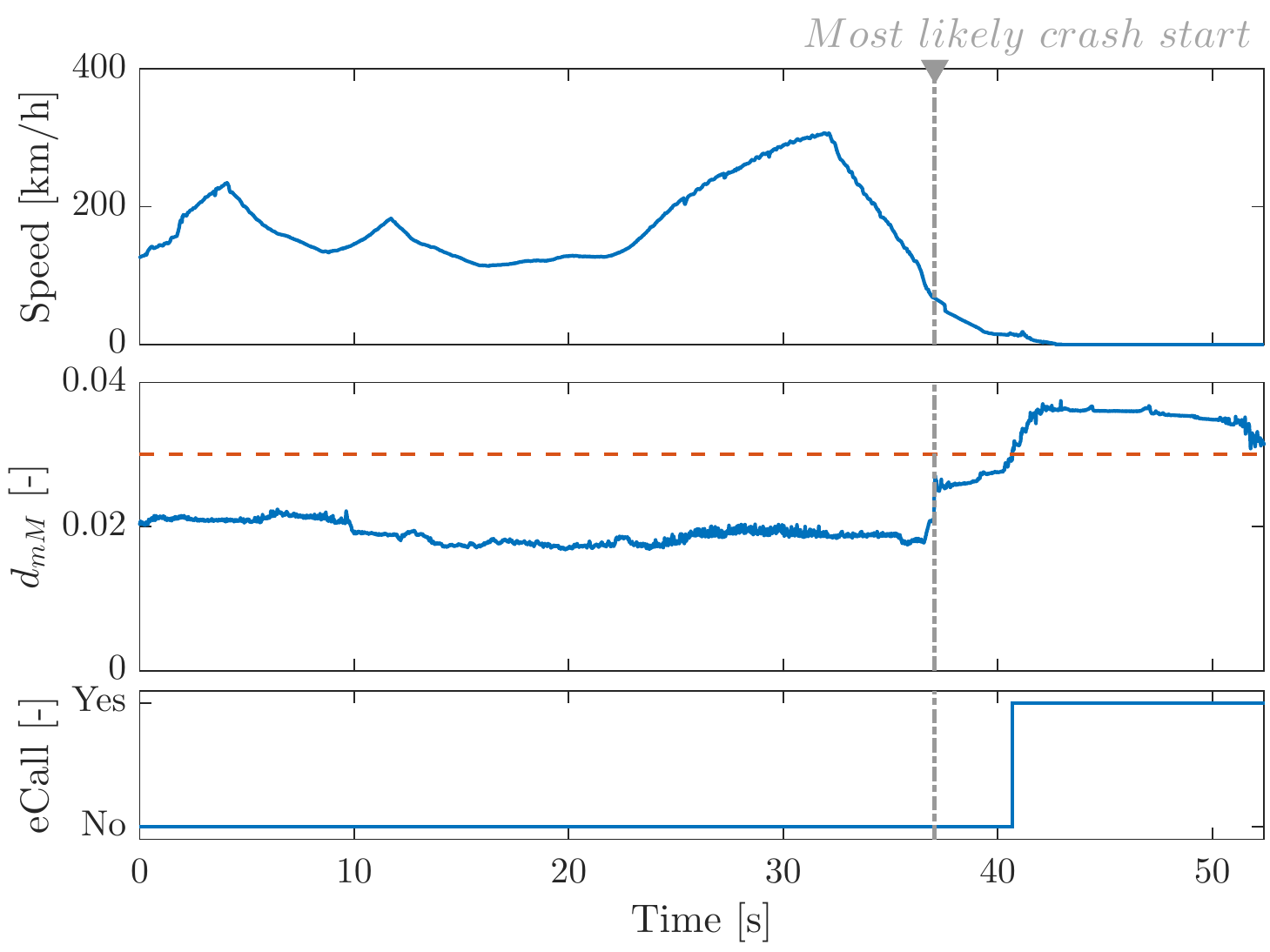}
	\caption{An example of the cepstrum-based detection on the \textit{Front lowside I} crash. The algorithm triggers the \textit{eCall} in $3.64$ seconds after the suspected start of the crash dynamics.}
	\label{fig:cepstrum_nominal_6}
\end{figure}

\begin{figure}[thpb]
	\centering
	\includegraphics[width=0.42\textwidth]{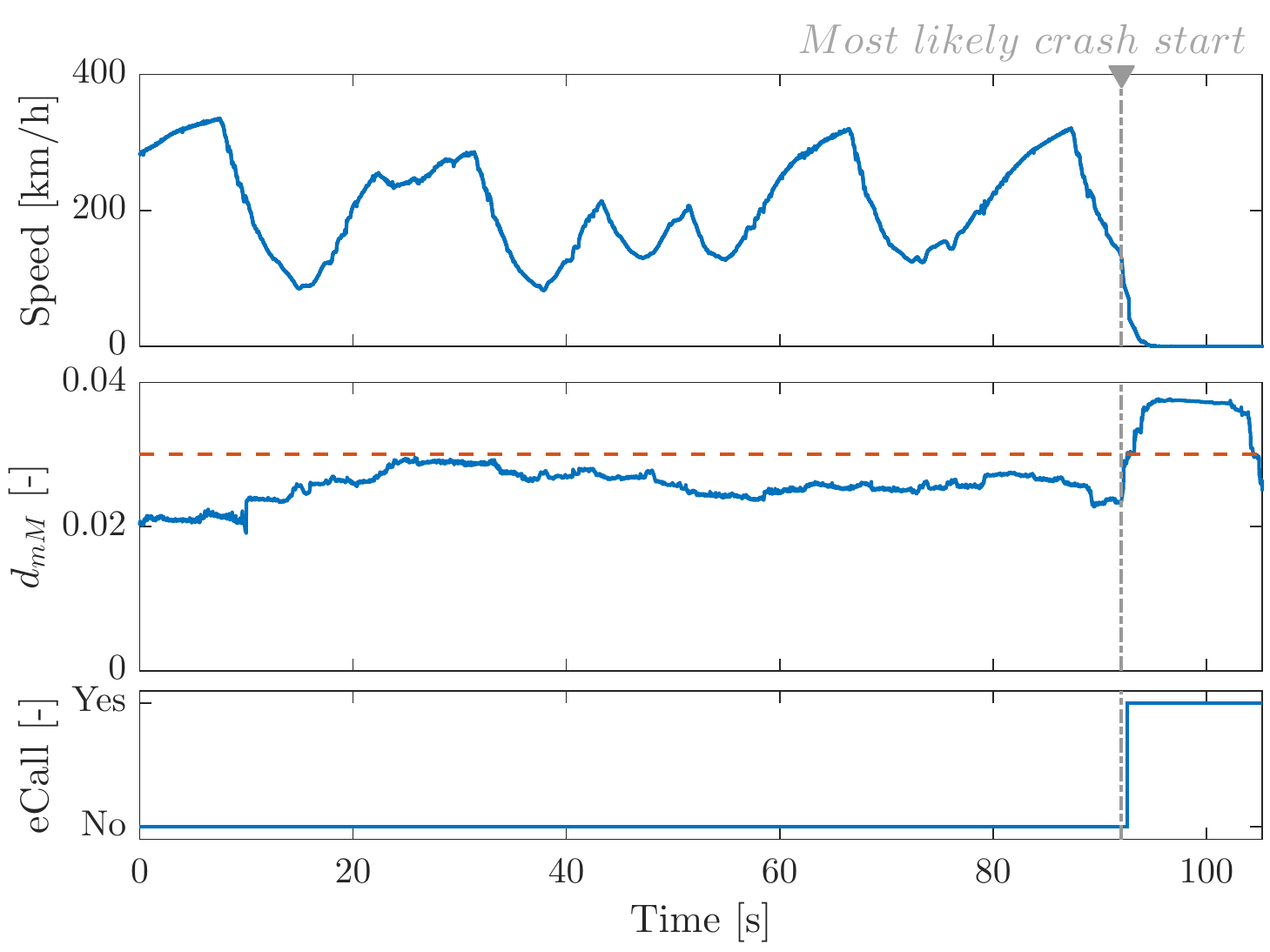}
	\caption{An example of the cepstrum-based detection on the \textit{Cornering lowside} crash.}
	\label{fig:cepstrum_nominal_4}
\end{figure}

To calibrate the threshold $\gamma_{cep}$, (\ref{martin_distance_one_class_application}) is computed before and during the crash event for the seven datasets (Fig. \ref{fig:sensitivity_threshold_cepstrum}): the maximum value of $d_{cep}$ reaches a peak of $0.027$ during the normal ride; instead, in the crash dynamics, the peak distance is always kept above $0.031$. Setting $\gamma_{cep} = 0.029$ makes possible to detect all the crashes: Fig. \ref{fig:cepstrum_nominal_6} and Fig. \ref{fig:cepstrum_nominal_4} show two examples of cepstrum-based detection for the \textit{Front lowside I} and \textit{Cornering lowside} crashes, respectively.
Beside the prompt crash detection, its consistency should be appreciated: the \textit{eCall} flag rises and keeps high for a significant amount of time, which is indirectly linked to the window length $w$, making the proposed cepstrum-based crash detection approach more appealing than the other methods.

Finally, to assess the robustness of the method with respect to the false positive triggers, the algorithm is tested with the proposed tuning against more than eighty hours of tests, not used in the calibration phase. The algorithm does not trigger any emergency call for the entire validation set, though different driving styles are involved.

\section{Concluding remarks}\label{sec:conclusions}
In the proposed contribution, a one-class, cepstrum-based algorithm was proposed to detect motorcycle accidents. Contrarily to the other works in literature, the algorithm triggers an emergency call by exploiting the signals dynamics through a multivariate cepstral analysis. Experimental results on seven real crashes and more than eighty hours of test favorably witnessed the effectiveness of the method.



\IEEEpeerreviewmaketitle
\bibliographystyle{ieeetr}
\bibliography{eCall_one_class_cepstrum}

\end{document}